\documentstyle[preprint,aps,axodraw]{revtex}
\tightenlines

\newcommand{\reseteqnum}{\setcounter{equation}{0}}
\newcommand{\nn}{\nonumber}

\newcommand{\wt}[1]{\widetilde{#1}}
\newcommand{\eqn}[1]{(\ref{#1})}
\newcommand{\p}{\partial}
\newcommand{\pslash}{p\kern-1ex /}
\newcommand{\kslash}{k\kern-1ex /}
\newcommand{\dslash}{\p\kern-1.2ex /}
\newcommand{\Aslash}{A\kern-1.2ex /}
\newcommand{\bpsi}{{\overline{\psi}}}

\newcommand{\bxi}{{\overline{\xi}}}
\newcommand{\tr}{{\rm tr}}
\newcommand{\vev}[1]{\left\langle #1 \right\rangle}

\begin{document}


\title{
\vspace{-3.0cm}
\begin{flushright}  
{\normalsize hep-lat/9906026}\\
{\normalsize UTHEP-406}\\
\end{flushright}
One loop calculation of SUSY Ward-Takahashi identity on lattice with
Wilson fermion}

\author{Yusuke Taniguchi}
\address{Institute of Physics, University of Tsukuba, Tsukuba,
 Ibaraki 305-8571, Japan\\}

\date{\today}
\maketitle

\begin{abstract}
One loop correction to the SUSY Ward-Takahashi identity is calculated on 
 lattice with Wilson fermion.
The supersymmetry on lattice is broken explicitly by the gluino mass
 and the lattice artifact.
We should fine tune parameters in the theory to the point given by the
 additive mass correction in order to eliminate the breaking effect of
 lattice artifact.
It is shown that the additive mass correction appearing from the SUSY 
 Ward-Takahashi identity coincide with that from the axial $U(1)_R$
 symmetry as was suggested by Curci and Veneziano.
Two important symmetries of the super Yang-Mills theory can be recovered
 simultaneously in the continuum with a single fine tuning.

Operator mixing of the supercurrent are also investigated.
We find that the supercurrent mixes only with a gauge invariant current
 $T_\mu$ which is related to the gamma-trace anomaly.

\end{abstract}

\pacs{11.15Ha, 11.30Pb, 11.30Rd, 12.38Bx, 12.38Gc}

\narrowtext

\reseteqnum
\section{Introduction}

There has been a great progress in nonperturbative understanding of the
low energy behavior of the $N=1$ supersymmetric QCD\cite{IS96}.
The analysis is based on the global symmetry and the holomorphy of the
superpotential and we can derive the nonperturbative form of
superpotential.
This is quite satisfactory when we investigate the vacuum structure of
the theory. However when it is required to understand the low energy
particle spectrum including excited states or an influence of the
K\"ahler potential this method is insufficient and some other
nonperturbative method is required.

The lattice regularization when applied to a supersymmetric theory
breaks its supersymmetry explicitly.
This is mainly caused by the artifact of the lattice regularization
itself and the fermion problem on lattice.
However ability of the lattice field theory to perform the path integral 
nonperturbatively with Monte-Carlo method is so fascinating that several 
efforts has been made to resolve the difficulty of SUSY on lattice.
These attempts are classified into two types.
One is to realize a SUSY on the lattice which corresponds to the ordinary
supersymmetry in the continuum limit\cite{AK98,Bietenholtz98,SU98}.
Although this method is beautiful in construction and can extract
peculiar feature of the model due to supersymmetry before taking the
continuum limit in principle , it is applicable only to the free
Wess-Zumino model up to now.
In order to treat the $N=1$ supersymmetric Yang-Mills (SYM) theory,
whose component fields (gluon and gluino) are forced to stay at
different places (links and sites) on lattice to keep the gauge
symmetry, we need the second method.
In this method we do not persist in the supersymmetry and
discretize the theory straightforwardly making use of the well known
actions on lattice.
The SUSY is recovered only in the continuum\cite{CV87}.
This restoration of supersymmetry is not automatic and the
discussion in Ref.~\cite{CV87} is as follows.

The SYM theory has two important global symmetries in the continuum.
One is supersymmetry and the other is axial $U(1)_R$ symmetry which is
broken by the anomaly.
Both the symmetries are broken explicitly when the theory is regularized
on lattice with the Wilson plaquette action for gluon and the Wilson
fermion action for gluino.
The source of this symmetry breaking is classified into the introduction
of gluino mass which cannot be forbidden in the Wilson fermion and the
lattice artifacts of discretization.
This explicit breaking effect of the lattice artifacts is given by
irrelevant operators in the Ward-Takahashi (WT) identity and vanishes in
the continuum at tree level.
However when a quantum correction comes into play this term usually
produces an additive corrections.
It is required to fine tune several parameters of the theory in order to 
recover the symmetry in the continuum.
Although this fine tuning should be performed independently for each
symmetries, we have only one parameter (gluino mass) in $N=1$ SYM which
we can freely tune.
It is discussed by Curci and Veneziano~\cite{CV87} that the both
symmetries are restored simultaneously with a single fine tuning of the
gluino mass to the chiral $U(1)_R$ symmetric point.
Several Monte-Carlo studies of SYM theory have been done along this line
\cite{Montvay95,DG96,KM96,DGHV97,DESY98,DESY99}
to reproduce the prediction of the low energy effective
theory\cite{VY82,FGS97}.

In this paper we formulate the $N=1$ SYM theory on lattice with the
Wilson plaquette and the Wilson fermion action according to
Ref.~\cite{CV87}.
We calculate the one loop correction to the Ward-Takahashi 
identity of both the SUSY and $U(1)_R$ symmetry perturbatively in the
gauge variant Green function.
It is shown that the additive mass correction appearing from the SUSY 
WT identity coincide with that from the axial $U(1)_R$
symmetry as was suggested.
This means that the both symmetries of the super Yang-Mills theory can
be recovered simultaneously in the continuum with a single fine tuning
of the gluino mass. 
We also investigate the mixing behavior of the supercurrent with the
on-shell condition for gluino momentum and mass.
The supercurrent mixes with the gauge invariant operator $T_\mu$ as was
predicted in Ref.~\cite{CV87}.
This current is related to the gamma-trace anomaly of the supercurrent.
An extra mixing with gauge variant operators occurs too.
This is because we used gauge variant Green function in our
calculation. 
However these extra mixings vanish by setting the renormalized gluino
mass to zero together with the on-shell condition.

This paper is organized as follows.
In Sec.~\ref{sec:model} we introduce the lattice SYM action and the
Feynman rules relevant for the one loop calculation.
In Sec.~\ref{sec:WT_id} we define the super transformation on lattice
and give the concerning SUSY WT identity.
The WT identity for $U(1)_R$ symmetry is also given in this section.
Sec.~\ref{sec:oneloop-axial} and \ref{sec:oneloop-SUSY} are devoted to
the calculation of quantum correction at one loop level for the axial
and SUSY Ward-Takahashi identity.
Our conclusion is summarized in Sec.~\ref{sec:concl}.

The physical quantities are expressed in lattice units 
and the lattice spacing $a$ is suppressed unless necessary. 
We take SU($N_c$) gauge group with the gauge coupling $g$,
the generator $T^a$ and the structure constant $f^{abc}$.
The normalization is given as $\tr(T^aT^b)=\frac{1}{2}\delta^{ab}$.

\reseteqnum
\section{Action and Feynman rule}
\label{sec:model}

The SYM theory is given as a minimally gauge coupled
massless adjoint Majorana fermion system in the continuum.
In this paper we adopt the following lattice regularization
procedure\cite{CV87}.
The gauge part is given by a standard four dimensional Wilson plaquette
action,
\begin{eqnarray}
S_{\rm gluon} &=&
\sum_n \sum_{\mu \nu}
-\frac{1}{g^2} {\rm Re} \; \tr
\left( U_{n, \mu} U_{n+\hat{\mu}, \nu}
U_{n+\hat{\nu}, \mu}^\dagger U_{n, \nu}^\dagger \right).
\end{eqnarray}
The gluino part is given by the Wilson fermion,
\begin{eqnarray}
S_{\rm gluino} &=&
\sum_n \tr \biggl[
 \frac{1}{2}\bpsi(n) \left(-r+\gamma_\mu\right)
 U_\mu(n) \psi(n+\mu) U_\mu^\dagger(n)
\nn\\&&\qquad
+\frac{1}{2}\bpsi(n+\hat{\mu}) \left(-r-\gamma_\mu\right)
 U_\mu^\dagger(n) \psi(n) U_\mu(n)
+(M+4r) \bpsi(n) \psi(n)
\biggr]
\nn\\&=&
\sum_n \tr \biggl[
 \bpsi(n) \left(-r+\gamma_\mu\right)
 U_\mu(n) \psi(n+\mu) U_\mu^\dagger(n)
+(M+4r) \bpsi(n) \psi(n)
\biggr],
\label{eqn:action}
\end{eqnarray}
where gluino filed $\psi=\psi^a T^a$ is the adjoint representation of
the gauge group and satisfies the Majorana condition,
\begin{eqnarray}
\psi = \psi^C = C \bpsi^T,
\quad
\bpsi = \bpsi^C = \psi^T \left(-C^{-1}\right).
\end{eqnarray}
The charge conjugation matrix is given as $C=\gamma_0\gamma_2$.
We used this condition in the second equality of \eqn{eqn:action}.
Our $\gamma$ matrix convention is as follows:
\begin{eqnarray}
&&
\gamma_i
= \pmatrix{
  0        & -i \sigma^i \cr
i \sigma^i & 0           \cr},
\quad
\gamma_4
= \pmatrix{
0 & 1 \cr
1 & 0 \cr},
\quad
\gamma_5 \equiv \gamma_1 \gamma_2 \gamma_3 \gamma_4
=\pmatrix{
1 &  0 \cr
0 & -1 \cr
} ,
\\&&
\sigma_{\mu \nu} = \frac{1}{2} \left[ \gamma_\mu , \gamma_\nu \right],
\\&&
\epsilon_{1234} = 1.
\end{eqnarray}

Weak coupling perturbation theory is developed by writing the link
variable as
$U_{x,\mu}=e^{igA_\mu(x+\hat{\mu}/2)}$
and expanding it in terms of gauge coupling $g$.
We adopt a covariant gauge fixing with a gauge parameter $\alpha$
defined by
\begin{eqnarray}
S_{\rm GF} &=&
 \sum_n \frac{1}{2\alpha}
\left( \nabla_\mu A_\mu^a \left(n+\frac{1}{2}\hat{\mu}\right) \right)^2.
\end{eqnarray}
We set $\alpha=1$ in this paper.
The ghosts do not contribute to our present calculation at one loop level.
The gluon propagator can be written as
\begin{eqnarray}
G_{\mu \nu}^{ab} (p)
=\frac{1}{4\sin^2 p/2}
\left[\delta_{\mu \nu}
-(1-\alpha)\frac{4 \sin {p}_\mu/2 \sin {p}_\nu/2}{4 \sin^2 p/2}
\right] \delta_{ab},
\end{eqnarray}
where $\sin^2 p/2 = \sum_\mu \sin^2 p_\mu/2$.

The free gluino propagator is the same as that of the Dirac fermion 
on lattice,
\begin{eqnarray}
S_{\rm F}^{ab}(p) \equiv \vev{\psi^a(p)\bpsi^b(-p)} =
\frac{-i \sum_{\mu}\gamma_\mu \sin p_\mu + W(p)}
{\sum_{\mu}\sin^2 p_\mu + W(p)^2}\delta_{ab},
\end{eqnarray}
where
\begin{eqnarray}
W(p) = M+r\sum_\mu \left(1-\cos p_\mu\right).
\end{eqnarray}
We set the Wilson parameter $r=1$ in this paper.
A peculiar feature of the Majorana fermion is that the propagators which
connect two $\psi$'s or two $\bpsi$'s give non zero
contribution
\begin{eqnarray}
\vev{\psi^a(p)\psi^b(-p)} = S_{\rm F}^{ab}(p) \left(-C\right),
\quad
\vev{\bpsi^a(p)\bpsi^b(-p)} = C^{-1} S_{\rm F}^{ab}(p).
\end{eqnarray}

\begin{figure}
\begin{center}\begin{picture}(200,100)(0,0)
\Text(-15,50)[r]{$V_{1\mu}^{ab,c}(k,p)$}
\ArrowLine(100,70)(10,90)
\Text(10,90)[r]{$\bpsi^a(k)$}
\Text(25,90)[lb]{$\rightarrow k$}
\ArrowLine(190,90)(100,70)
\Text(193,90)[l]{$\psi^b(p)$}
\Text(175,90)[rb]{$p \leftarrow$}
\Gluon(100,10)(100,70){5}{5}
\Text(115,45)[l]{$l$}\LongArrow(115,25)(115,35)
\Text(85,45)[r]{$A^c_\mu(l)$}
\Vertex(100,70){3}
\end{picture}
\end{center}
\begin{center}\begin{picture}(200,100)(0,0)
\Text(-15,50)[r]{$V_{2\mu\nu}^{ab,cd}(k,p)$}
\ArrowLine(100,70)(10,90)
\Text(10,90)[r]{$\bpsi^a(k)$}
\Text(25,90)[lb]{$\rightarrow k$}
\ArrowLine(190,90)(100,70)
\Text(193,90)[l]{$\psi^b(p)$}
\Text(175,90)[rb]{$p \leftarrow$}
\Gluon(50,10)(100,70){5}{5}
\Text(65,55)[l]{$l_1$}\LongArrow(40,30)(60,50)
\Text(45,15)[r]{$A^{c}_\mu(l_1)$}
\Gluon(150,10)(100,70){5}{5}
\Text(135,55)[r]{$l_2$}\LongArrow(160,30)(140,50)
\Text(155,15)[l]{$A^{d}_\nu(l_2)$}
\Vertex(100,70){3}
\end{picture}
\end{center}
\begin{center}\begin{picture}(200,100)(0,0)
\Text(-15,50)[r]{$G_{3\mu\nu\rho}^{abc}(k,l,p)\ $}
\Gluon(100,70)(10,90){5}{5}
\Text(10,90)[r]{$A_\rho^c(p)$}
\Text(25,100)[lb]{$\rightarrow p$}
\Gluon(190,90)(100,70){5}{5}
\Text(193,90)[l]{$A_\nu^b(l)$}
\Text(175,100)[rb]{$l \leftarrow$}
\Gluon(100,10)(100,70){5}{5}
\Text(115,45)[l]{$k$}\LongArrow(115,25)(115,35)
\Text(85,5)[]{$A^a_\mu(k)$}
\Vertex(100,70){3}
\end{picture}
\caption[]{Quark-gluon vertices needed for our one-loop calculations.}
\label{fig:rule}
\end{center}
\end{figure}
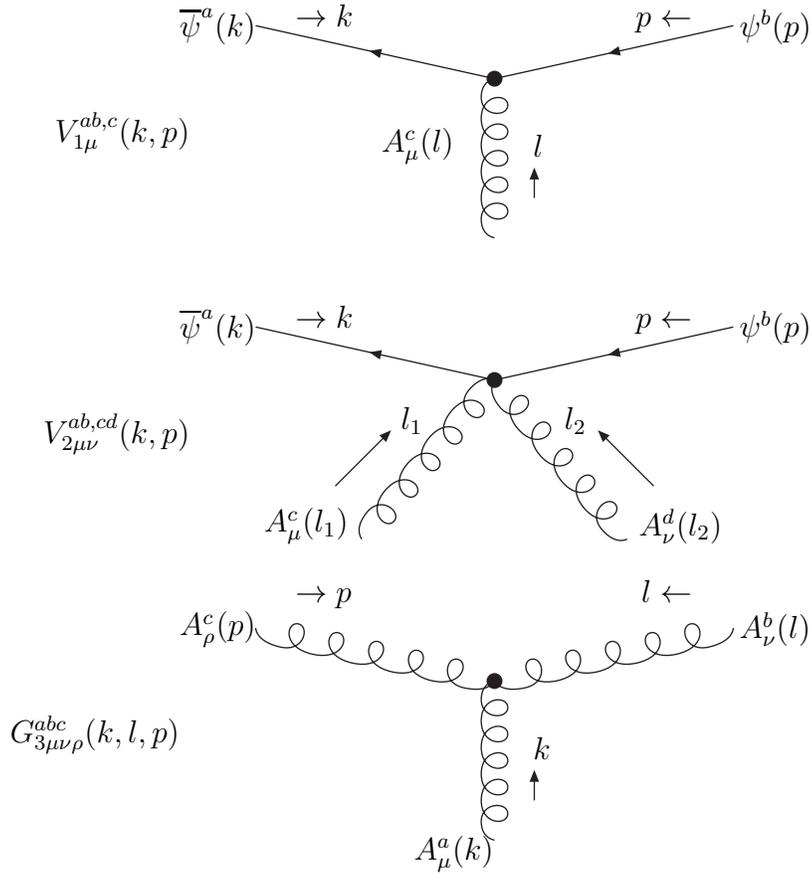

In order 
to calculate the one loop correction to the SUSY WT identity 
we need two kinds of gluon-gluino interaction vertex
\begin{eqnarray}
V_{1\mu}^{ab,c}(k,p)
&=& -\frac{1}{2} g f^{abc}
 \{ \gamma_\mu \cos \frac{1}{2}(-k_\mu + p_\mu)
-i r \sin \frac{1}{2}(-k_\mu + p_\mu) \},
\\
V_{2\mu\nu}^{ab,cd}(k,p) &=&
\frac{1}{8} g^2
\left(f^{ace}f^{ebd} + f^{ade}f^{ebc}\right)
\{ i \gamma_\mu \sin \frac{1}{2}(-k_\mu + p_\mu)
-r \cos \frac{1}{2} (-k_\mu + p_\mu) \}\delta_{\mu\nu}
\nn\\
\end{eqnarray}
and three gluon self interaction vertex,
\begin{eqnarray}
G_{3\mu\nu\rho}^{abc}(k,l,p) &=&
\frac{1}{3} ig f^{abc}
\biggl(
\delta_{\nu\rho}
\cos\frac{k_\nu}{2} \sin\frac{1}{2}(p-l)_\mu
\nn\\&&
+\delta_{\rho\mu}
\cos\frac{l_\rho}{2} \sin\frac{1}{2}(k-p)_\nu
+\delta_{\mu\nu}
\cos\frac{p_\mu}{2} \sin\frac{1}{2}(l-k)_\rho
\biggr).
\end{eqnarray}
Our assignments of momentum and color factors for the vertices are
depicted in Fig.~\ref{fig:rule}.

\reseteqnum
\section{Axial and SUSY Ward-Takahashi identity on lattice}
\label{sec:WT_id}

The $U(1)_R$ transformation is given as an axial rotation of the gluino
field,
\begin{eqnarray}
\delta \psi = i \alpha \gamma_5 \psi,
\quad
\delta \bpsi = i \alpha \bpsi \gamma_5
\end{eqnarray}
with a rotation parameter $\alpha$.
The corresponding axial Ward-Takahashi identity is given by
\begin{eqnarray}
\vev{\left(\nabla_\mu j_{5\mu}(n)\right) {\cal O}}
= 2M \vev{D_A(n) {\cal O}} + \vev{X_A(n) {\cal O}}
-\vev{\frac{\delta{\cal O}}{\delta \alpha(n)}},
\end{eqnarray}
where the axial current $j_{5\mu}$ and the symmetry breaking terms
$D_A$, $X_A$ are
\begin{eqnarray}
j_{5\mu}(n) &=&
\tr \left(
\bpsi(n) \gamma_\mu\gamma_5 U_\mu(n) \psi(n+\mu) U_\mu^\dagger(n)\right),
\\
D_A(n) &=& \tr \left(\bpsi(n) \gamma_5 \psi(n)\right),
\\
X_A(n) &=&
-r \sum_\mu \tr \Bigl(
 \bpsi(n)\gamma_5 U_\mu(n) \psi(n+\mu) U_\mu^\dagger(n)
+\bpsi(n)\gamma_5 U_\mu^\dagger(n-\mu) \psi(n-\mu) U_\mu(n-\mu)
\nn\\&&\qquad
-2\bpsi(n) \gamma_5 \psi(n)
\Bigr).
\end{eqnarray}
$\nabla_\mu$ is a backward derivative,
${\cal O}$ is some operator and $\alpha(n)$ is a localized
transformation parameter.
The trace is taken for the color indices only.

On the other hand there are several choices for the definition of the
supertransformation on lattice.
The restriction is only to recover the proper form in the continuum
limit.
Adding to this condition we require the supertransformation to commute
with the parity transformation on lattice as in the continuum,
\begin{eqnarray}
&&
{\cal P} \psi(\vec{x},t) {\cal P}^{-1} = \gamma_0 \psi(-\vec{x},t),
\quad
{\cal P} \bpsi(\vec{x},t) {\cal P}^{-1} = \bpsi(-\vec{x},t) \gamma_0,
\\&&
{\cal P} U_0(\vec{x},t) {\cal P}^{-1} = U_0(-\vec{x},t),
\quad
{\cal P} U_k(\vec{x},t) {\cal P}^{-1} = U_k^\dagger(-\vec{x}-\hat{k},t).
\end{eqnarray}
In this paper we adopt the following definition to satisfy the above
conditions, 
\begin{eqnarray}
&&
\delta_\xi U_\mu(n) =
 ig \bxi \gamma_\mu \frac{1}{2}
\left(\psi(n) U_\mu(n) + U_\mu(n) \psi(n+\mu) \right),
\\&&
\delta_\xi U_\mu^\dagger(n) =
 -ig \bxi \gamma_\mu \frac{1}{2}
\left(U_\mu^\dagger(n) \psi(n) + \psi(n+\mu) U_\mu^\dagger(n)\right),
\\&&
\delta_\xi \psi(n) =
-\frac{1}{2} \sigma_{\mu\nu} \xi P_{\mu\nu}(n),
\\&&
\delta_\xi \bpsi(n) =
\frac{1}{2} \bxi \sigma_{\mu\nu} P_{\mu\nu}(n),
\end{eqnarray}
where $\xi$, $\bxi$ are fermionic transformation parameter satisfying
the Majorana condition.
For the field strength $P_{\mu\nu}$ we employ the definition with clover
plaquette,
\begin{eqnarray}
&&
P_{\mu\nu}(n) = \frac{1}{4} \sum_{i=1}^4 \frac{1}{2ig}
\left(U_{i \mu\nu}(n)-U_{i \mu\nu}^\dagger(n)\right),
\\&&
U_{1 \mu\nu} (n) = U_{\mu}(n) U_{\nu}(n+\hat{\mu})
 U^\dagger_{\mu}(n+\hat{\nu}) U^\dagger_{\nu}(n),
\\&&
U_{2 \mu\nu} (n) = U_{\nu}(n) U^\dagger_{\mu}(n-\hat{\mu}+\hat{\nu})
 U^\dagger_{\nu}(n-\hat{\mu}) U_{\mu}(n-\hat{\mu}),
\\&&
U_{3 \mu\nu} (n) =
 U^\dagger_{\mu}(n-\hat{\mu}) U^\dagger_{\nu}(n-\hat{\mu}-\hat{\nu})
 U_{\mu}(n-\hat{\mu}-\hat{\nu}) U_{\nu}(n-\hat{\nu}),
\\&&
U_{4 \mu\nu} (n) = U^\dagger_{\nu}(n-\hat{\nu}) U_{\mu}(n-\hat{\nu})
 U_{\nu}(n+\hat{\mu}-\hat{\nu}) U^\dagger_{\mu}(n).
\end{eqnarray}
This definition is slightly different from the original one\cite{CV87}.

By transforming the vacuum expectation value of some operator
\begin{eqnarray}
\vev{{\cal O}} = \int dU d\psi {\cal O}e^{-S_{\rm gluon}-S_{\rm gluino}}
\end{eqnarray}
with a localized transformation parameter we find the SUSY WT identity
on lattice,
\begin{eqnarray}
\vev{\left(\nabla_\mu S_\mu(n)\right) {\cal O}}
&=&
M \vev{D_S(n) {\cal O}} + \vev{X_S(n) {\cal O}}
- \vev{\frac{\delta {\cal O}}{\delta \bxi(n)}},
\end{eqnarray}
where the supercurrent $S_\mu$ and the gluino mass term $D_S$ become
\begin{eqnarray}
S_\mu(n) &=&
-\frac{1}{2} \sum_{\rho\sigma} \sigma_{\rho\sigma} \gamma_\mu
 \tr \biggl(
 P_{\rho\sigma}(n) U_\mu(n)\psi(n+\hat{\mu})U_\mu^\dagger(n)
+P_{\rho\sigma}(n+\mu) U_\mu^\dagger(n)\psi(n)U_\mu(n)
\biggr),
\nn\\
\\
D_S(n) &=&
\sum_{\rho\sigma} \sigma_{\rho\sigma}
 \tr \left(P_{\rho\sigma}(n) \psi(n)\right).
\end{eqnarray}
The explicit SUSY breaking term $X_S$ is given by a sum of four terms
\begin{eqnarray}
X_S(n) &=& X_S^{(1)}(n)+X_S^{(2)}(n)+X_S^{(3)}(n)+X_S^{(4)}(n)
\end{eqnarray}
with
\begin{eqnarray}
X_S^{(1)}(n) &=&
\sum_{\mu\rho\sigma} r \sigma_{\rho\sigma} \tr \biggl[
P_{\rho\sigma}(n)
\Bigl(
 \psi(n)
-\frac{1}{2} \left(U_\mu(n) \psi(n+\mu) U_\mu^\dagger(n)\right)
\nn\\&&\quad
-\frac{1}{2} \left(U_\mu^\dagger(n-\mu)\right)\psi(n-\mu)U_\mu(n-\mu)
\Bigr)
\biggr],
\\
X_S^{(2)}(n) &=&
\sum_{\mu\nu\rho\sigma} \frac{1}{2} \epsilon_{\mu\nu\rho\sigma}
 \gamma_\nu \gamma_5 \tr \Bigl(
 U_\mu^\dagger(n)\psi(n)U_\mu(n) P_{\rho\sigma}(n+\mu)
\nn\\&&
-U_\mu(n-\mu)\psi(n)U_\mu^\dagger(n-\mu) P_{\rho\sigma}(n-\mu)
\Bigr),
\\
X_S^{(3)}(n) &=&
\sum_{\mu\nu} \gamma_\nu \tr \Bigl(
 U_\mu(n-\mu)\psi(n)U_\mu^\dagger(n-\mu) H_{\mu\nu}(n-\mu)
+U_\mu^\dagger(n)\psi(n)U_\mu(n) H_{\mu\nu}(n+\mu)
\nn\\&&\quad
-2\psi(n) H_{\mu\nu}(n) \Bigr),
\\
X_S^{(4)}(n) &=&
-\frac{1}{2}ig \sum_\mu
\biggl[
\left(\gamma_\mu \psi^a(n)\right)
 \bpsi^b(n) \left(-r+\gamma_\mu\right) \psi^c(n+\mu)
\tr \left(\left[T^a,T^b\right] U_\mu(n) T^c U_\mu^\dagger(n)\right)
\nn\\&&
+\left(\gamma_\mu \psi^a(n)\right)
 \bpsi^b(n) \left(r+\gamma_\mu\right) \psi^c(n-\mu)
\tr \left(\left[T^a,T^b\right] U_\mu^\dagger(n-\mu) T^c U_\mu(n-\mu)\right)
\biggr],
\nn\\
\end{eqnarray}
where $H_{\mu\nu}$ is given by a subtraction between the clover leafs,
\begin{eqnarray}
H_{\mu\nu}(n) &=& \frac{1}{8ig}
\Bigl(
 \left(U_{1, \mu\nu}(n)-U_{1, \mu\nu}^\dagger(n)\right)
+\left(U_{4, \mu\nu}(n)-U_{4, \mu\nu}^\dagger(n)\right)
\nn\\&&\quad
-\left(U_{2, \mu\nu}(n)-U_{2, \mu\nu}^\dagger(n)\right)
-\left(U_{3, \mu\nu}(n)-U_{3, \mu\nu}^\dagger(n)\right)
\Bigr).
\end{eqnarray}
Here we notice that $X_S^{(2)}$, $X_S^{(3)}$ come from the pure gauge
part and $X_S^{(1)}$, $X_S^{(4)}$ are originated from the gluino action.
Especially $X_S^{(1)}$ is given by transforming the gluino field in the
Wilson term.

At tree level the breaking terms due to the Wilson fermion $X_S^{(1)}$,
$X_S^{(4)}$ represent the ${\cal O}(a)$ irrelevant operator and those
from the plaquette action $X_S^{(2)}$, $X_S^{(3)}$ represent the
${\cal O}(a^2)$ operator.
The SUSY WT identity is recovered in the continuum with the gluino mass
set to zero since all the irrelevant operators vanish and the lattice
supercurrent $S_\mu(n)$ gives the continuum form
\begin{eqnarray}
S_\mu(n) \to -\sum_{\rho\sigma} \sigma_{\rho\sigma}\gamma_\mu
 \tr \left(F_{\rho\sigma} \psi\right).
\end{eqnarray}
However at one loop order every term gives the finite contribution and
the symmetry restoration becomes nontrivial.

Derivation of the Feynman rules for the supercurrent and breaking terms
is straightforward but tedious task and the results are very
complicated.
We omitted them in this paper.

\reseteqnum
\section{One loop correction to axial Ward-Takahashi identity}
\label{sec:oneloop-axial}

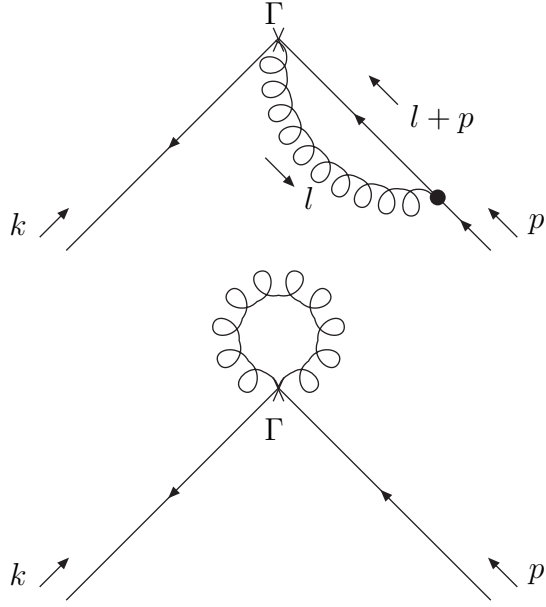
\begin{figure}
\begin{center}\begin{picture}(200,100)(0,0)
\ArrowLine(180,10)(160,30)
\ArrowLine(160,30)(100,90)
\ArrowLine(100,90)(40,30)
\ArrowLine(40,30)(20,10)
\Gluon(40,30)(160,30){5}{5}
\Vertex(40,30){3}
\Vertex(160,30){3}
\Line(98,85)(102,94)
\Line(98,94)(102,85)
\Text(150,60)[l]{$l+p$}\LongArrow(145,65)(135,75)
\Text(50,75)[l]{$l-k$}\LongArrow(60,70)(50,60)
\Text(95,20)[l]{$l$}\LongArrow(100,20)(110,20)
\Text(95,100)[l]{$\Gamma$}
\Text(195,20)[l]{$p$}\LongArrow(190,15)(180,25)
\Text(5,20)[r]{$k$}\LongArrow(10,15)(20,25)
\end{picture}
\end{center}
\begin{center}
\begin{picture}(200,100)(0,0)
\ArrowLine(180,10)(100,90)
\ArrowLine(100,90)(40,30)
\ArrowLine(40,30)(20,10)
\GlueArc(50,80)(50,-100,10){-5}{10}
\Vertex(40,30){3}
\Line(98,85)(102,94)
\Line(98,94)(102,85)
\Text(90,30)[l]{$l$}\LongArrow(95,35)(105,45)
\Text(50,75)[l]{$l-k$}\LongArrow(60,70)(50,60)
\Text(95,100)[l]{$\Gamma$}
\Text(195,20)[l]{$p$}\LongArrow(190,15)(180,25)
\Text(5,20)[r]{$k$}\LongArrow(10,15)(20,25)
\end{picture}
\end{center}
\begin{center}
\begin{picture}(200,100)(0,0)
\ArrowLine(180,10)(160,30)
\ArrowLine(160,30)(100,90)
\ArrowLine(100,90)(20,10)
\GlueArc(150,80)(52,170,280){-5}{10}
\Vertex(160,30){3}
\Line(98,85)(102,94)
\Line(98,94)(102,85)
\Text(150,60)[l]{$l+p$}\LongArrow(145,65)(135,75)
\Text(110,30)[l]{$l$}\LongArrow(95,45)(105,35)
\Text(95,100)[l]{$\Gamma$}
\Text(195,20)[l]{$p$}\LongArrow(190,15)(180,25)
\Text(5,20)[r]{$k$}\LongArrow(10,15)(20,25)
\end{picture}
\end{center}
\begin{center}
\begin{picture}(200,120)(0,0)
\ArrowLine(180,10)(100,90)
\ArrowLine(100,90)(20,10)
\GlueArc(100,110)(20,-90,270){-5}{10}
\Line(98,85)(102,94)
\Line(98,94)(102,85)
\Text(95,75)[l]{$\Gamma$}
\Text(195,20)[l]{$p$}\LongArrow(190,15)(180,25)
\Text(5,20)[r]{$k$}\LongArrow(10,15)(20,25)
\end{picture}
\end{center}
\caption[]{One loop diagrams which contribute to the operator vertex
 correction in the axial Ward-Takahashi identity.
 $\Gamma=j_{5\mu}, D_A, X_A$ for the first diagram and
 $\Gamma=j_{5\mu}, X_A$ for the remaining three.}
\label{fig:axial}
\end{figure}

In this section we calculate the one loop correction to the axial
Ward-Takahashi identity.
We consider the WT identity with 
${\cal O}=\psi(y)\bpsi(z)$:
\begin{eqnarray}
0&=&
\vev{\left(\nabla_\mu j_{5\mu}(n)\right)\psi(y)\bpsi(z)}
-2M \vev{D_A(n)\psi(y)\bpsi(z)}
-\vev{X_A(n)\psi(y)\bpsi(z)}
\nn\\&&
+\delta_{n,y} \gamma_5 \vev{\psi(n) \bpsi(z)}
+\delta_{n,z} \vev{\psi(y) \bpsi(n)} \gamma_5.
\label{eqn:axial-WTid}
\end{eqnarray}
We calculate the quantum correction to each Green functions in the
identity.
Although our fermion is Majorana the one loop correction becomes the
same as that for the Dirac fermion system \cite{BMMRT} except for the
color factor.
One loop contributions to the Green functions
 $\vev{\left(\nabla_\mu j_{5\mu}\right)\psi\bpsi}$ and
 $\vev{X_A\psi\bpsi}$
are given by the four diagrams in Fig.~\ref{fig:axial}.
One loop level full Green functions become
\begin{eqnarray}
&&
\vev{\left(\nabla_\mu j_{5\mu}\right)\psi(k)\bpsi(p)}^{\rm full} =
 \frac{Z_2}{-i\kslash+Z_m^{-1}M}i(k+p)_\mu T_A\gamma_\mu\gamma_5
 \frac{Z_2}{i\pslash+Z_m^{-1}M},
\\&&
\vev{X_A\psi(k)\bpsi(p)}^{\rm full} =
 \frac{Z_2}{-i\kslash+Z_m^{-1}M}
 \left(i(k+p)_\mu X_a\gamma_\mu\gamma_5+(X_m M +X_0) \gamma_5\right)
 \frac{Z_2}{i\pslash+Z_m^{-1}M}.
\end{eqnarray}
One loop correction to $\vev{D_A \psi\bpsi}$ is given by the
first diagram in Fig.~\ref{fig:axial},
\begin{eqnarray}
\vev{D_A\psi(k)\bpsi(p)}^{\rm full} &=&
 \frac{Z_2}{-i\kslash+Z_m^{-1}M}T_P\gamma_5
 \frac{Z_2}{i\pslash+Z_m^{-1}M}.
\end{eqnarray}

The vertex corrections $T_A$ and $T_P$ depend on the gluino
external momentum and mass and they have infra-red divergence at
$k=p=M=0$ in general.
In order to regularize the infra-red singularity we adopt the following
subtraction scheme,
\begin{eqnarray}
T_A(k,p,M) = \left(T_A(k,p,M)-T_A^{\rm cont.}(k,p,M)\right)
 +T_A^{\rm cont.}(k,p,M),
\label{eqn:subtraction}
\end{eqnarray}
where $T_A^{\rm cont.}$ is a vertex correction given by integrating the
continuum Feynman rule with lattice loop momentum between $-\pi/a$ and
$\pi/a$. 
Since $T_A^{\rm cont.}$ has the same IR singularity, the IR divergence
is subtracted in the first term of \eqn{eqn:subtraction}.
We evaluate it by a Taylor expansion around $(k,p,M)=(0,0,0)$.
The second term should be calculated analytically keeping the IR
regulator finite.
In this paper we introduce a gluon mass $\lambda$ into the gluon
propagator inside the loop as an IR regularization
\footnote{
Strictly speaking this IR regularization procedure cannot be applied to 
all the quantum corrections in the SUSY WT identity because there is a
gluon propagator in the external line and its wave function
renormalization is required.
The IR divergence of the gluino and ghost loops in the gluon
polarization cannot be regularized.
However we proceed our calculation with this scheme since we are
only interested in the one loop vertex correction whose IR divergence is 
treatable in this method.}.
We can evaluate $T_A^{\rm cont.}$ quite simply with a Taylor expansion
in terms of $(k,p,M)$ keeping $\lambda$ finite,
\begin{eqnarray}
T_A^{\rm cont.}(k,p,M;\lambda) = T_A^{\rm cont.}(0;\lambda)+
k_\mu \frac{\p T_A^{\rm cont.}(0;\lambda)}{\p k_\mu} +\cdots,
\end{eqnarray}
where $T_A^{\rm cont.}(0;\lambda)$ contribute to the renormalization of
the operator and the remaining terms are ${\cal O}(a)$ errors.

In this scheme the vertex corrections are given as follows
\begin{eqnarray}
&&
T_A = 1+\frac{g^2}{16\pi^2} N_c
\left[ -\log\frac{(\lambda a)^2}{\pi^2} -6.977 \right],
\\&&
T_P = 1+\frac{g^2}{16\pi^2} N_c
\left[ -4\log\frac{(\lambda a)^2}{\pi^2} + 2.585 \right],
\\&&
X_a =\frac{g^2}{16\pi^2} N_c (8.664),
\\&&
X_m =\frac{g^2}{16\pi^2} N_c (-19.285),
\\&&
X_0 =\frac{g^2}{16\pi^2} N_c (102.8694).
\end{eqnarray}
The gluino wave function and mass renormalization factors are evaluated
by the quantum correction to the gluino propagator,
\begin{eqnarray}
&&
Z_2 = 1+\frac{g^2}{16\pi^2} N_c
\left[ \log\frac{(\lambda a)^2}{\pi^2} + 15.641 \right],
\label{eqn:gluino-wave}
\\&&
Z_m = 1+\frac{g^2}{16\pi^2} N_c
\left[ 3 \log\frac{(\lambda a)^2}{\pi^2} -8.584 \right].
\end{eqnarray}
The renormalization of operators is given to keep the proper form of the 
axial WT identity,
\begin{eqnarray}
G_A&=&
\vev{\left(\nabla_\mu j_{5\mu}\right)\psi(k)\bpsi(p)}^{\rm full}
-2M \vev{D_A\psi(k)\bpsi(p)}^{\rm full}
-\vev{X_A\psi(p)\bpsi(p)}^{\rm full}
\nn\\&=&
 \frac{Z_2}{-i\kslash+Z_m^{-1}M}
\biggl[
i(k+p)_\mu \left(T_A-X_a\right)\gamma_\mu\gamma_5
-2 \left(M\left(T_P+\frac{X_m}{2}\right)+\frac{X_0}{2}\right)\gamma_5
\biggr]
 \frac{Z_2}{i\pslash+Z_m^{-1}M}
\nn\\
\\&=&
Z_2 \left(
Z_A \vev{\left(\p_\mu j_{5\mu}\right)\psi(k)\bpsi(p)}_R
-2M_R \vev{D_A\psi(k)\bpsi(p)}_R
\right),
\end{eqnarray}
where suffix $R$ means renormalized quantity.
The gluino mass is renormalized as
\begin{eqnarray}
M_R = Z_m^{-1}\left(M - \Sigma_0\right),
\end{eqnarray}
where $\Sigma_0$ is an additive mass correction
\begin{eqnarray}
\Sigma_0 = -\frac{X_0}{2}= \frac{g^2}{16\pi^2}N_c(-51.4347),
\label{eqn:additive-axial}
\end{eqnarray}
which gives the critical hopping parameter corresponding to the chiral
symmetric point
\begin{eqnarray}
K_c = \frac{1}{8}\left(1-\frac{\Sigma_0}{4}\right).
\end{eqnarray}
The multiplicative mass renormalization factor evaluated from the WT
identity agrees with that from the gluino propagator
\begin{eqnarray}
Z_m^{-1} = Z_2 \left(T_P+\frac{X_m}{2}\right)
=1-\frac{g^2}{16\pi^2} N_c
 \left[ 3 \log\frac{(\lambda a)^2}{\pi^2} -8.584 \right].
\end{eqnarray}
The renormalization factors of the axial current becomes unity in order
to keep the axial WT identity
\begin{eqnarray}
Z_A = Z_2\left(T_A-X_a\right) = 1.000.
\end{eqnarray}
Numerical errors of the finite parts in this section are in the last
digit written.

\reseteqnum
\section{One loop correction to SUSY Ward-Takahashi identity}
\label{sec:oneloop-SUSY}

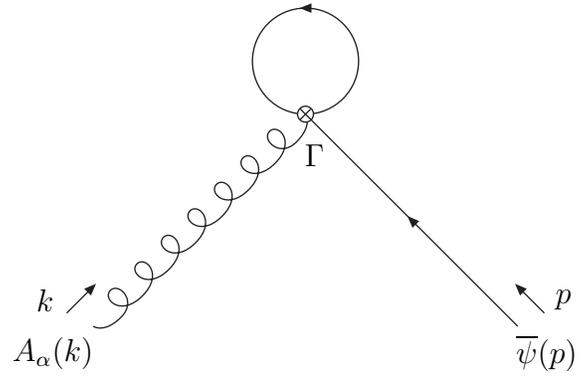
\begin{figure}
\begin{center}\begin{picture}(200,90)(0,-10)
\ArrowArc(80,37)(25,33,213)
\ArrowArc(80,37)(25,213,393)
\Vertex(59,24){2}
\Gluon(59,24)(0,-5){-5}{4}
\Text(-5,-10)[r]{$A_\alpha(k)$}
\ArrowLine(190,-5)(100,50)
\Text(195,-10)[l]{$\bpsi(p)$}
\BCirc(100,50){4}
\Line(97,47)(103,53)
\Line(97,53)(102,47)
\Text(110,60)[rb]{$\Gamma$}
\Text(15,25)[r]{$k$}
\LongArrow(10,13)(22,20)
\Text(160,35)[l]{$p$}
\LongArrow(157,25)(147,32)
\Text(75,80)[l]{$l$}
\LongArrow(70,75)(60,67)
\Text(110,-5)[r]{$l+k$}
\LongArrow(90,3)(101,11)
\end{picture}\end{center}
\vspace*{0.5cm}
\begin{center}
\begin{picture}(200,120)(0,0)
\ArrowLine(180,10)(100,90)
\Gluon(100,90)(20,10){5}{7}
\ArrowArc(100,110)(20,-90,270)
\BCirc(100,90){3}
\Line(98,88)(102,92)
\Line(98,92)(102,88)
\Text(20,0)[r]{$A_\alpha(k)$}
\Text(180,0)[l]{$\bpsi(p)$}
\Text(100,75)[l]{$\Gamma$}
\Text(195,20)[l]{$p$}\LongArrow(190,15)(180,25)
\Text(5,20)[r]{$k$}\LongArrow(10,15)(20,25)
\end{picture}
\end{center}
\caption[]{One loop diagrams which contribute to the explicit SUSY
 breaking term; $\Gamma=X_S^{(4)}$.}
\label{fig:SUSY-ex}
\end{figure}

We consider the following SUSY WT identity with
${\cal O}=A_\alpha(y)\psi(z)$:
\begin{eqnarray}
0&=&
\vev{\left(\nabla_\mu S_\mu(n)\right)A_\alpha(y)\bpsi(z)}
-M\vev{D_S(n)A_\alpha(y)\bpsi(z)}
- \vev{X_S(n)A_\alpha(y)\bpsi(z)}
\nn\\&&
+\delta_{n,y} \gamma_\alpha\vev{\psi(y)\bpsi(z)}
+\delta_{n,z} \frac{1}{2}\sigma_{\rho\sigma}
 \vev{A_\alpha(y)P_{\rho\sigma}(z)}.
\label{eqn:SUSY-WTid}
\end{eqnarray}
Although these Green functions are gauge variant we can derive
gauge independent quantities such as additive mass correction correctly.
One loop corrections to the Green functions
$\vev{\left(\nabla_\mu S_\mu\right)A_\alpha\bpsi}$,
$\vev{D_SA_\alpha\bpsi}$ and $\vev{X_S^{(i)}A_\alpha\bpsi}$ with
$i=1,2,3$ are given by the diagrams in Fig.~\ref{fig:SUSY}.
The one loop correction to the explicit breaking term with $i=4$ is
given by the two diagrams in Fig.~\ref{fig:SUSY-ex}.

As was discussed in the previous section we introduce the gluon mass
$\lambda$ here again in order to settle the IR singularity at vanishing
external momentum and gluino mass.
All the IR divergence in the one loop vertex correction
appears from the first four diagrams of Fig.~\ref{fig:SUSY}
and can be regularized in this scheme.
The fermion loops in Fig.~\ref{fig:SUSY-ex} do not have any IR
divergence.

We calculate the one loop contribution imposing the on-shell condition
to the external gluino momentum,
\begin{eqnarray}
\Gamma (i\pslash+M_R)=0,
\end{eqnarray}
where $\Gamma$ represents some operator vertex.
This is applicable to the bare mass inside the one loop correction by
making use of the relation $g^2M=g^2M_R+{\cal O}(g^4)$.
The one loop correction to the Green functions
$\vev{\left(\nabla_\mu S_\mu\right)A_\alpha\bpsi}$ and
$\vev{D_S A_\alpha\bpsi}$ are
\begin{eqnarray}
\vev{\left(\nabla_\mu S_\mu\right)A_\alpha(k)\bpsi(p)}_1 &=&
\frac{1}{k^2} (k_\mu+p_\mu) \Bigl(
 k_\nu\sigma_{\nu\alpha}\gamma_\mu T_S^{(S)}
+\left(\delta_{\mu\alpha}\kslash-k_\mu \gamma_\alpha\right) T_T^{(S)}
\Bigr) \frac{1}{i\pslash+M},
\\
\vev{D_S A_\alpha(k)\bpsi(p)}_1 &=&
\frac{1}{k^2}\left(
iM k_\mu\sigma_{\mu\alpha} T_D^{(D)}
+p_\alpha\pslash T_{GP}^{(D)}
+p^2 \gamma_\alpha T_{GM}^{(D)}
\right) \frac{1}{i\pslash+M},
\end{eqnarray}
where
\begin{eqnarray}
&&
T_S^{(S)} = \frac{g^2}{16\pi^2}N_c
 \left(-27.874(1)+\frac{2\pi^2}{N_c^2}\right),
\\&&
T_T^{(S)} = \frac{g^2}{16\pi^2}N_c
 \left(6.372(1)\right),
\\&&
T_D^{(D)} = \frac{g^2}{16\pi^2}N_c
\left(-3\ln\frac{(\lambda a)^2}{\pi^2}
 -23.6453(7)+\frac{2\pi^2}{N_c^2}\right),
\\&&
T_{GP}^{(D)} = \frac{g^2}{16\pi^2}N_c \left(1.000(1)\right),
\\&&
T_{GM}^{(D)} = \frac{g^2}{16\pi^2}N_c \left(-1.000(1)\right).
\end{eqnarray}
The finite term proportional to $1/N_c$ emerges from the last tadpole
diagram in Fig.~\ref{fig:SUSY}.
In the above the finite part of the loop correction is evaluated by
performing the loop integrals with the Monte Carlo routine VEGAS in
double precision. 
We employ $20$ sets of $10^6$ points for integration.
Errors are estimated from variation of integrated values over the sets.  
We eliminated the terms when their coefficients become smaller than the
numerical errors.
Here we remind that the logarithmic divergence appears only in the term
concerning to the mass renormalization.

The one loop contribution to the explicit breaking term
$\vev{X_S A_\alpha\bpsi}$ is given by
\begin{eqnarray}
\vev{X_S A_\alpha(k)\bpsi(p)}_1 &=&
\frac{1}{k^2} \biggl[
(k+p)_\mu
 \left(k_\nu \sigma_{\nu\alpha} \gamma_\mu T_S^{(X)}
+\left(\delta_{\mu\alpha}\kslash-k_\mu\gamma_\alpha\right) T_T^{(X)}
\right)
\nn\\&&
+ik_\mu \sigma_{\mu\alpha} \left(T_D^{\rm add.}+M T_D^{(X)}\right)
\nn\\&&
+k_\alpha \pslash T_{GK}^{(X)}
+p^2 \gamma_\alpha T_{GM}^{(X)}
\biggr] \frac{1}{i\pslash+M},
\end{eqnarray}
where
\begin{eqnarray}
&&
T_S^{(X)} = \frac{g^2}{16\pi^2}N_c \left(-7.775(1)\right),
\\&&
T_T^{(X)} = \frac{g^2}{16\pi^2}N_c \left(3.3716(7)\right),
\\&&
T_D^{\rm add.} = \frac{g^2}{16\pi^2}N_c (51.4345(5)),
\label{eqn:additive-susy}
\\&&
T_D^{(X)} = \frac{g^2}{16\pi^2}N_c \left(11.130(1)\right),
\\&&
T_{GK}^{(X)} = \frac{g^2}{16\pi^2}N_c \left(-2.000(1)\right),
\\&&
T_{GM}^{(X)} = \frac{g^2}{16\pi^2}N_c \left(-1.842(2)\right).
\end{eqnarray}
Here we should notice that the one loop correction to the SUSY explicit
breaking term $X_S$ produces an additive mass correction given by
\eqn{eqn:additive-susy}.
This additive correction coincide with that from the axial
Ward-Takahashi identity \eqn{eqn:additive-axial} within a numerical
error.
This fact confirms the prediction of Ref.~\cite{CV87}.
We have a comment on the origin of this additive correction.
The SUSY explicit breaking term $X_S$ can be classified into four parts
$X_S^{(i)}$ .
$X_S^{(1)}$ is given by supertransforming the gluino fields of the
Wilson term, $X_S^{(4)}$ appears from the gluino action by transforming
the link variable in the covariant derivative.
$X_S^{(2)}$, $X_S^{(3)}$ are originated from the gluon plaquette action.
Since the additive mass in the axial WT identity \eqn{eqn:additive-axial} 
is due to the axial symmetry breaking term given by rotating the Wilson
term, it might have been expected that the additive correction in the SUSY
WT identity comes only from $X_S^{(1)}$, $X_S^{(4)}$ which are directly
related to the Wilson term.
However $X_S^{(1)}$, $X_S^{(4)}$ produces only $80\%$ of
$T_D^{\rm add.}$ and remaining $20\%$ is a contribution from $X_S^{(2)}$,
$X_S^{(3)}$.
Contributions from each $X_S^{(i)}$ is given in table \ref{tbl:each}.
The Wilson parameter dependence of $T_D^{\rm add.}$ is given
nontrivially inside the diagram multiplying $X_S^{(2)}$, $X_S^{(3)}$
with the Wilson parameter in the gluino propagator and the interaction
vertex. 
We depicted the $r$ dependence of $\Sigma_0$ and $T_D^{\rm add.}$ in
table \ref{tbl:wilson-r}, which are in good agreement for every $r$
within a numerical error.

Summing up all the contributions we can investigate the mixing behavior.
\begin{eqnarray}
G_S^{(1)} &=& 
 \vev{\left(\nabla_\mu S_\mu\right)A_\alpha(k)\bpsi(p)}_1
-M\vev{D_S A_\alpha(k)\bpsi(p)}_1
-\vev{X_S A_\alpha(k)\bpsi(p)}_1
\nn\\&=&
\frac{1}{k^2} \biggl[
(k_\mu+p_\mu) \left(
 k_\nu \sigma_{\nu\alpha}\gamma_\mu T_S^{(1)}
+\left(\delta_{\mu\alpha}\kslash-k_\mu\gamma_\alpha\right) T_T^{(1)}
\right)
\nn\\&&\quad
-ik_\mu\sigma_{\mu\alpha} \left(M T_D^{(1)}+T_D^{\rm add.}\right)
\nn\\&&\quad
-k_\alpha\pslash T_{GK}
-p_\alpha\pslash T_{GP}
-p^2 \gamma_\alpha T_{GM}
\biggr] \frac{1}{i\pslash+M},
\label{eqn:mixing}
\end{eqnarray}
where
\begin{eqnarray}
T_S^{(1)} &=& T_S^{(S)}-T_S^{(X)}
=\frac{g^2}{16\pi^2}N_c \left(-20.099(2)+\frac{2\pi^2}{N_c^2}\right),
\\
T_T^{(1)} &=& T_T^{(S)}-T_T^{(X)}
=\frac{g^2}{16\pi^2}N_c \left(3.000(2)\right),
\\
T_D^{(1)} &=& T_D^{(D)}+T_D^{(X)}
=\frac{g^2}{16\pi^2}N_c
 \left(-3\ln\frac{(\lambda a)^2}{\pi^2}-12.515(2)
  +\frac{2\pi^2}{N_c^2}\right),
\\
T_{GK} &=& T_{GK}^{(X)} = \frac{g^2}{16\pi^2}N_c \left(-2.000(1)\right),
\\
T_{GP} &=& T_{GP}^{(D)} = \frac{g^2}{16\pi^2}N_c \left(1.000(1)\right),
\\
T_{GM} &=& T_{GM}^{(D)}+T_{GM}^{(X)}
 = \frac{g^2}{16\pi^2}N_c \left(-2.842(3)\right).
\end{eqnarray}

Now we consider the mixing property of the operators by making use of
the continuum form of the Green functions
\begin{eqnarray}
&&
\vev{(\p_\mu S_\mu) A_\alpha(k)\bpsi(p)} =
\frac{1}{k^2} (k_\mu+p_\mu) k_\nu \sigma_{\nu\alpha} \gamma_\mu
\frac{1}{i\pslash+M},
\\&&
\vev{(\p_\mu T_\mu) A_\alpha(k)\bpsi(p)} =
\frac{1}{k^2} (k_\mu+p_\mu)
\left(\delta_{\mu\alpha}\kslash-k_\mu\gamma_\alpha\right)
\frac{1}{i\pslash+M},
\\&&
\vev{D_S A_\alpha(k)\bpsi(p)} =
\frac{1}{k^2} M ik_\nu \sigma_{\nu\alpha} \frac{1}{i\pslash+M},
\\&&
\vev{\left(\p_\mu A_\mu\dslash\psi\right)A_\alpha(k)\bpsi(p)} =
-\frac{1}{k^2} k_\alpha\pslash \frac{1}{i\pslash+M},
\\&&
\vev{\left(A_\mu\p_\mu \dslash\psi\right)A_\alpha(k)\bpsi(p)} =
-\frac{1}{k^2} p_\alpha\pslash \frac{1}{i\pslash+M},
\\&&
\vev{\left(\Aslash \p_\mu\p_\mu\psi\right)A_\alpha(k)\bpsi(p)} =
-\frac{1}{k^2} p^2 \gamma_\alpha \frac{1}{i\pslash+M}.
\end{eqnarray}
Here we introduce a gauge invariant fermionic current with dimension
$7/2$
\begin{eqnarray}
T_\mu(n) = 2 \tr\left(P_{\mu\nu}(n)\gamma_\nu\psi(n)\right).
\end{eqnarray}
We can easily see that the first term in \eqn{eqn:mixing} with
$T_S^{(1)}$ contribute to the multiplicative normalization factor of the
supercurrent.
The second term with $T_T^{(1)}$ represents the mixing with a gauge
invariant current $T_\mu$ as was discussed in Ref.~\cite{CV87}.
Mixing with $T_\mu$ is also reported in the continuum theory with the
dimensional regularization\cite{AGS}.
This term is related to the gamma-trace anomaly corresponding to the
super conformal symmetry breaking.
Its coefficient $T_T^{(1)}$ is identical with the one loop level
$\beta$-function of the $N=1$ SYM,
\begin{eqnarray}
T_T^{(1)} = \frac{g^2}{16\pi^2}3N_c = -\frac{\beta_{\rm 1-loop}}{g}
\end{eqnarray}
as was required for the gamma-trace anomaly\cite{VY82}.
The third term gives the additive and multiplicative mass correction.
The remaining three terms are mixing with gauge variant operators.
This mixing is because we adopted gauge variant operator
${\cal O}=A_\alpha(y)\psi(z)$ in \eqn{eqn:SUSY-WTid} and fixed gauge in
perturbative calculation.
However these extra mixings disappear if we impose on-shell condition to 
the gluino momentum and set the renormalized gluino mass to zero
\begin{eqnarray}
i\pslash = -M_R =0.
\end{eqnarray}

Hereafter we assume the existence of a consistent IR regularization
scheme with the gluon mass procedure for the gluon wave function.
If the gluon wave function renormalization factor $Z_3$ is derived in a
well defined manner the one loop level full Green function can be given
\begin{eqnarray}
G_S &=& 
 \vev{\left(\nabla_\mu S_\mu\right)A_\alpha(k)\bpsi(p)}^{\rm full}
-M\vev{D_S A_\alpha(y)\bpsi(z)}^{\rm full}
-\vev{X_S A_\alpha(k)\bpsi(p)}^{\rm full}
\nn\\&=&
Z_S \sqrt{Z_3 Z_2} \vev{\left(\nabla_\mu S_\mu\right)A_\alpha(k)\bpsi(p)}_R
\nn\\&&
-Z_D \sqrt{Z_3 Z_2} \left(M+T_D^{\rm add.}\right)
 \vev{D_S A_\alpha(k)\bpsi(p)}_R
\nn\\&&
+T_T^{(1)} \vev{\left(\nabla_\mu T_\mu\right)A_\alpha(k)\bpsi(p)}_R
\nn\\&&
+T_{GK} \vev{\left(\p_\mu A_\mu\dslash\psi\right)A_\alpha(k)\bpsi(p)}_R
\nn\\&&
+T_{GP} \vev{\left(A_\mu\p_\mu \dslash\psi\right)A_\alpha(k)\bpsi(p)}_R
\nn\\&&
+T_{GM} \vev{\left(\Aslash \p_\mu\p_\mu\psi\right)A_\alpha(k)\bpsi(p)}_R,
\end{eqnarray}
where $Z_2$ is the gluino wave function renormalization factor
\eqn{eqn:gluino-wave}.
The renormalization factors are given as
\begin{eqnarray}
Z_S = \sqrt{Z_3 Z_2}\left(1+T_S^{(1)}\right),
\quad
Z_D = \sqrt{Z_3 Z_2}\left(1+T_D^{(1)}\right).
\end{eqnarray}
The logarithmic divergences in the gluon and gluino wave function
renormalization factor are canceled with the multiplication $Z_3Z_2$ as
is easily confirmed in the continuum with dimensional regularization
and the supercurrent renormalization factor $Z_S$ remains finite.
The renormalization of the gluino mass is
\begin{eqnarray}
\wt{M}_R = Z_D \left(M+T_D^{\rm add.}\right),
\end{eqnarray}
which gives the same critical mass as in \eqn{eqn:additive-axial}.

\reseteqnum
\section{Conclusion}
\label{sec:concl}

In this article we regularize the supersymmetric Yang-Mills theory on
lattice with the Wilson plaquette action for gluon and the Wilson
fermion for gluino.
In this regularization the supersymmetry and axial $U(1)_R$ symmetry of
the continuum SYM theory is broken explicitly.
However both of the symmetries can be recovered in the continuum by fine
tuning the mass parameter.

In order to see this restoration process we calculated the one loop
correction to the SUSY Ward-Takahashi identity perturbatively.
It is shown that the additive mass correction (the critical mass)
given by the SUSY Ward-Takahashi identity coincides with that from the
axial WT identity.
This means the SUSY and the $U(1)_R$ symmetry can be restored
simultaneously in the continuum limit with a single fine tuning of the
gluino mass.
Since tuning to the chiral symmetric point is a well known subject in
computer simulation, there would have been no technical difficulty to
deal with the SYM system on lattice even nonperturbatively if the axial
$U(1)_R$ symmetry had no anomaly.
The chiral symmetric point cannot be given by the vanishing pion mass
for the anomalous $U(1)_R$ symmetry of SYM and an alternative method is
needed.
Application of the vacuum degeneracy of residual $Z_{2N_c}$ symmetry due 
to gluino condensation seems to be hopeful\cite{DESY98}.

The supercurrent on lattice mixes nontrivially with $T_\mu$.
If we can extract this gamma-trace anomaly part nonperturbatively from
the SUSY WT identity we may be able to evaluate the exact
$\beta$-function of the $N=1$ SYM theory.

A peculiar point in our calculation is that the SUSY breaking term
given by supertransforming the plaquette action also contribute to the
additive mass correction.
Therefore when we consider to use the domain-wall fermion as a gluino
part, it is nontrivial to see disappearance of the additive mass
correction.
Because the domain-wall fermion system contains the Wilson term in its
action before integrating out the unphysical heavy modes.
The negative unity Wilson parameter remains in the gluino-gluon
interaction vertex.
This is a fascinating future problem.

\section*{Acknowledgments}
I greatly appreciate the valuable discussions with
S.~Aoki, T.~Izubuchi, T.~Kobayashi, Y.~Sato and A.~Ukawa.
Their comments were precise and helped me very much.
This work is supported in part by the Grants-in-Aid for Scientific
Research from the Ministry of Education, Science and Culture (No.2373).
Y.~T. is supported by Japan Society for Promotion of Science.

\newcommand{\J}[4]{{#1} {\bf #2} (19#3) #4}
\newcommand{\AP}{Ann.~Phys.}
\newcommand{\CMP}{Commun.~Math.~Phys.}
\newcommand{\IJMP}{Int.~J.~Mod.~Phys.}
\newcommand{\MPL}{Mod.~Phys.~Lett.}
\newcommand{\NP}{Nucl.~Phys.}
\newcommand{\NPSup}{Nucl.~Phys.~B (Proc.~Suppl.)}
\newcommand{\PL}{Phys.~Lett.}
\newcommand{\PR}{Phys.~Rev.}
\newcommand{\PRL}{Phys.~Rev.~Lett.}
\newcommand{\PTP}{Prog. Theor. Phys.}
\newcommand{\Suppl}{Prog. Theor. Phys. Suppl.}

\begin{table}[bht]
\caption{Contribution to additive mass correction from each $X_S^{(i)}$
 at $r=1$.
  $80\%$ of the contribution is given by $X_S^{(1)}$ and $X_S^{(4)}$.}
\label{tbl:each}
\begin{center}
\begin{tabular}{l|lllll}
$X_S^{i}$ & $X_S^{(1)}$ & $X_S^{(2)}$ & $X_S^{(3)}$ & $X_S^{(4)}$ \\
\hline 
$T_D^{\rm add.}(i)$ & $42.2083(5)$ & $7.32191(2)$
 & $2.85525(4)$ & $-0.95093(2)$ \\
\end{tabular}
\end{center}
\end{table}
\begin{table}[bht]
\caption{Wilson parameter dependence of the additive mass correction
 from axial and SUSY WT identity. Both results are in good agreement for 
 every $r$ within a numerical error.}
\label{tbl:wilson-r}
\begin{center}
\begin{tabular}{l|lllllll}
$r$ & $0$ & $0.2$ & $0.4$ & $0.6$ & $0.8$ & $1.0$ \\
\hline 
$T_D^{\rm add.}$ & $0$ & $19.793(8)$ & $30.707(8)$ & $38.286(7)$ &
 $44.964(8)$ & $51.4345(5)$ \\
$\Sigma_0$       & $0$ & $19.791(1)$ & $30.695(2)$ & $38.283(3)$ &
 $44.960(4)$ & $51.4346(1)$ \\
\end{tabular}
\end{center}
\end{table}

\begin{figure}
\vspace*{0.5cm}
\begin{center}\begin{picture}(200,80)(0,-20)
\Gluon(40,12)(100,80){5}{5}
\Vertex(40,12){2}
\Gluon(12,-20)(40,12){5}{2}
\Text(0,-18)[r]{$A_\alpha(k)$}
\Gluon(40,12)(160,12){5}{5}
\ArrowLine(160,12)(100,80)
\Vertex(160,12){2}
\ArrowLine(188,-20)(160,12)
\Text(195,-18)[l]{$\bpsi(p)$}
\BCirc(100,80){4}
\Line(97,77)(103,83)
\Line(97,83)(102,77)
\Text(105,87)[rb]{$\Gamma$}
\Text(5,8)[l]{$k$}
\LongArrow(11,-3)(20,7)
\Text(195,8)[r]{$p$}
\LongArrow(184,-3)(176,7)
\Text(100,0)[c]{$l\rightarrow$}
\LongArrow(63,57)(54,47)
\Text(53,57)[r]{$l-k$}
\LongArrow(144,47)(135,57)
\Text(150,55)[l]{$l+p$}
\end{picture}\end{center}
\vspace*{0.5cm}
\begin{center}\begin{picture}(200,80)(0,-20)
\Gluon(160,12)(100,80){-5}{5}
\Vertex(160,12){2}
\ArrowLine(188,-20)(160,12)
\Text(0,-18)[r]{$A_\alpha(k)$}
\ArrowLine(40,12)(100,80)
\Vertex(40,12){2}
\Gluon(12,-20)(40,12){5}{2}
\Text(195,-18)[l]{$\bpsi(p)$}
\ArrowLine(160,12)(40,12)
\BCirc(100,80){4}
\Line(97,77)(103,83)
\Line(97,83)(102,77)
\Text(105,87)[rb]{$\Gamma$}
\Text(5,8)[l]{$k$}
\LongArrow(12,-3)(21,7)
\Text(195,8)[r]{$p$}
\LongArrow(184,-3)(176,7)
\Text(155,55)[l]{$l-p$}
\LongArrow(138,57)(147,47)
\Text(50,57)[r]{$l+k$}
\LongArrow(56,47)(65,57)
\Text(100,0)[c]{$\leftarrow l$}
\end{picture}\end{center}
\vspace*{0.5cm}
\begin{center}\begin{picture}(200,90)(0,-20)
\GlueArc(80,37)(25,33,213){-5}{6}
\GlueArc(80,37)(25,213,393){-5}{6}
\Vertex(59,24){2}
\Gluon(59,24)(0,-5){-5}{4}
\Text(-5,-10)[r]{$A_\alpha(k)$}
\ArrowLine(190,-5)(100,50)
\Text(195,-10)[l]{$\bpsi(p)$}
\BCirc(100,50){4}
\Line(97,47)(103,53)
\Line(97,53)(102,47)
\Text(110,60)[rb]{$\Gamma$}
\Text(15,25)[r]{$k$}
\LongArrow(10,13)(22,20)
\Text(160,35)[l]{$p$}
\LongArrow(157,25)(147,32)
\Text(75,80)[l]{$l$}
\LongArrow(70,75)(60,67)
\Text(110,-5)[r]{$l+k$}
\LongArrow(90,3)(101,11)
\end{picture}\end{center}
\vspace*{0.5cm}
\begin{center}
\begin{picture}(200,100)(0,0)
\ArrowLine(180,10)(160,30)
\ArrowLine(160,30)(100,90)
\Gluon(100,90)(0,15){5}{7}
\GlueArc(150,80)(52,170,280){-4.5}{7}
\Vertex(160,30){3}
\BCirc(100,90){3}
\Line(98,88)(102,92)
\Line(98,92)(102,88)
\Text(0,5)[r]{$A_\alpha(k)$}
\Text(180,0)[l]{$\bpsi(p)$}
\Text(150,60)[l]{$l+p$}\LongArrow(145,65)(135,75)
\Text(110,30)[l]{$l$}\LongArrow(95,45)(105,35)
\Text(95,100)[l]{$\Gamma$}
\Text(195,20)[l]{$p$}\LongArrow(190,15)(180,25)
\Text(5,35)[r]{$k$}\LongArrow(10,35)(20,45)
\end{picture}
\end{center}
\vspace*{0.5cm}
\begin{center}
\begin{picture}(200,100)(0,0)
\ArrowLine(180,10)(140,50)
\ArrowLine(140,50)(100,90)
\Gluon(140,49)(0,15){-5}{7}
\GlueArc(120,70)(30,-45,140){-4.5}{6}
\Vertex(140,50){3}
\BCirc(100,90){3}
\Line(98,88)(102,92)
\Line(98,92)(102,88)
\Text(0,5)[r]{$A_\alpha(k)$}
\Text(180,0)[l]{$\bpsi(p)$}
\Text(90,70)[r]{$l+k+p$}\LongArrow(105,68)(95,78)
\Text(170,68)[l]{$l$}\LongArrow(160,75)(160,60)
\Text(90,100)[l]{$\Gamma$}
\Text(195,20)[l]{$p$}\LongArrow(190,15)(180,25)
\Text(5,35)[r]{$k$}\LongArrow(10,35)(24,39)
\end{picture}
\end{center}
\vspace*{0.5cm}
\begin{center}
\begin{picture}(200,120)(0,0)
\ArrowLine(180,10)(100,90)
\Gluon(100,90)(20,10){5}{7}
\GlueArc(100,110)(20,-90,270){-5}{7}
\BCirc(100,90){3}
\Line(98,88)(102,92)
\Line(98,92)(102,88)
\Text(20,0)[r]{$A_\alpha(k)$}
\Text(180,0)[l]{$\bpsi(p)$}
\Text(100,75)[l]{$\Gamma$}
\Text(195,20)[l]{$p$}\LongArrow(190,15)(180,25)
\Text(5,20)[r]{$k$}\LongArrow(10,15)(20,25)
\end{picture}
\end{center}
\caption[]{One loop diagrams which contribute to the operator vertex
 correction in the SUSY WT identity. $\Gamma=S_\mu, D_S, X_S^{(1-3)}$.}
\label{fig:SUSY}
\end{figure}

\begin{references}

\bibitem{IS96} For a review, see:
K.~Intriligator and N.~Seiberg,
Nucl.~Phys. {\bf B45} (Proc.~Suppl.) (1996) 1
and references therein.

\bibitem{AK98} T.~ Aoyama and Y.~Kikukawa,
\J{\PR}{D59}{99}{054507}.

\bibitem{Bietenholtz98} W.~Bietenholz,
\J{\MPL}{A14}{99}{51}.

\bibitem{SU98} H.~So and N.~Ukita ,
hep-lat/9812002.

\bibitem{CV87} G.~Curci and G.~ Veneziano,
\J{\NP}{B292}{87}{555}.

\bibitem{Montvay95} I.~Montvay,
\J{\NP}{B466}{96}{259}.

\bibitem{DG96}
A.~Donini and M.~Guagnelli,
\J{\PL}{B383}{96}{301}.

\bibitem{KM96} G.~Koutsoubas and I.~Montvay,
\J{\PL}{B398}{97}{130}.

\bibitem{DGHV97} A.~Donini, M.~Guagnelli, P.~Hernandez and A.~Vladikas,
\J{\NP}{B523}{98}{529}.

\bibitem{DESY98} DESY-M\"unster Collaboration,
R.~Kirchner, S.~Luckmann, I.~Montvay, K.~Spanderen and J.~Westphalen,
\J{\NPSup}{73}{99}{828}; \J{\PL}{B446}{99}{209}.

\bibitem{DESY99} DESY-Munster Collaboration,
I.~Campos, A.~Feo, R.~Kirchner, S.~Luckmann, I.~Montvay, G.~M\"unster,
K.~Spanderen, J.~Westphalen,
hep-lat/9903014.

\bibitem{VY82} G.~Veneziano and S.~Yankielowicz,
\J{\PL}{B113}{82}{213}.

\bibitem{FGS97} G.R.~Farrar, G.~Gabadadze, M.~Schwetz,
\J{\PR}{D58}{98}{015009}.

\bibitem{BMMRT} M.~Bochicchio, L.~Maiani, G.~Martinelli, G.~Rossi and
 M.~Testa,
\J{\NP}{B262}{85}{331}.

\bibitem{AGS} L.~F.~Abbott, M.~T.~Grisaru and H.~J.~Schnitzer,
\J{\PR}{D16}{77}{2995}.

\end{references}
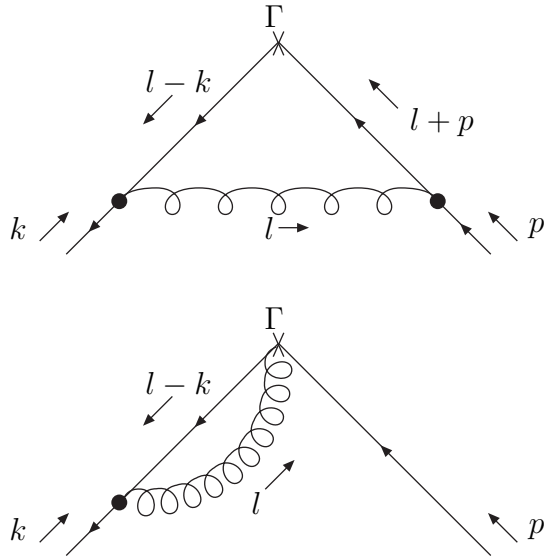
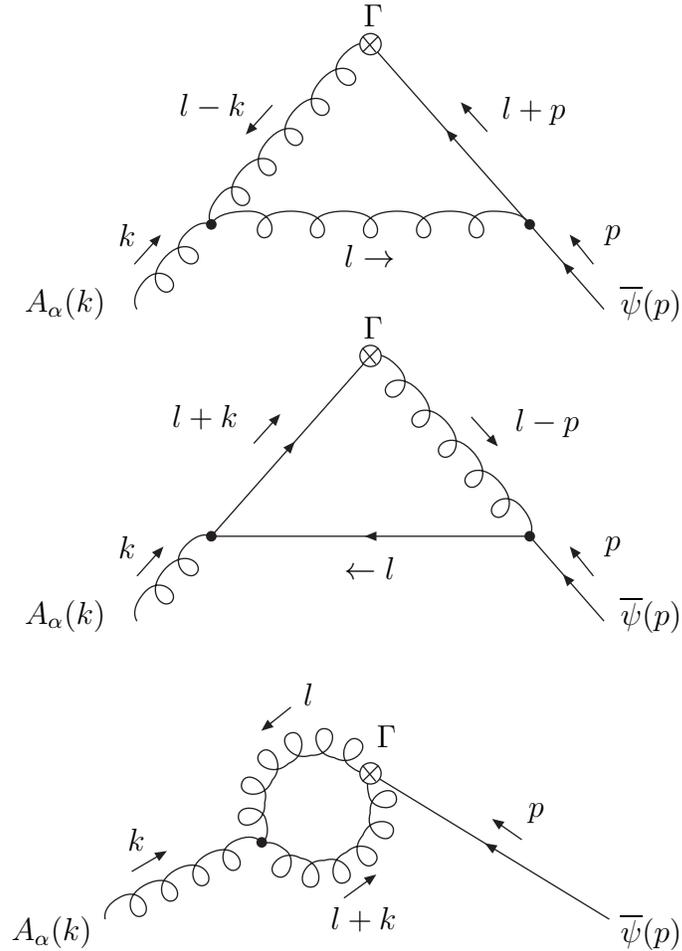
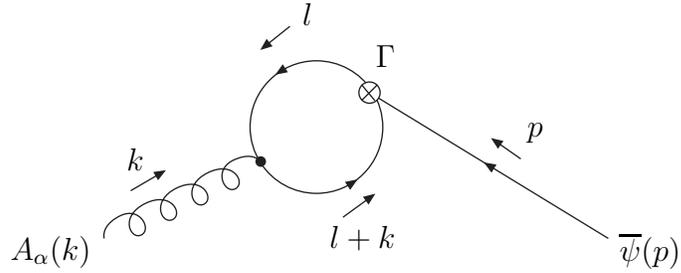
\end{document}